\documentclass[ALICE,manyauthors]{cernphprep}
\usepackage[comma,square,numbers,sort&compress]{natbib}
\usepackage{lineno}
\usepackage{xspace}
\usepackage{hyperref}
\usepackage{siunitx}
\usepackage{tikzducks}
\usepackage{tikzlings}
\usepackage[T1]{fontenc}
\usepackage{soul}
\usepackage{bm}
\usepackage{multirow}
\usepackage{amsmath}
\usepackage{bm} 

\DeclareSIUnit\clight{\text{\ensuremath{c}}}
\DeclareSIUnit\eVc{\eV/\clight} 
\DeclareSIUnit\keVc{\keV/\clight} 
\DeclareSIUnit\MeVc{\MeV/\clight} 
\DeclareSIUnit\MeVcc{\MeV/\clight\squared} 
\DeclareSIUnit\GeVc{\GeV/\clight} 
\DeclareSIUnit\GeVcc{\GeV/\clight\squared} 
\usepackage[Symbol]{upgreek}
\usepackage{xcolor}
\usepackage{verbatim}
\usepackage{subcaption}
\usepackage{amssymb}
\makeatletter

\newcommand{\Rmnum}[1]{\expandafter\@slowromancap\romannumeral #1@}
\makeatother

\usepackage{orcidlink}

\begin{document}

%

\newcommand{\pp}           {pp\xspace}
\newcommand{\pbpb}           {Pb--Pb\xspace}

\newcommand{\roots}        {\ensuremath{\sqrt{s}}\xspace}
\newcommand{\pt}           {\ensuremath{p_{\rm T}}\xspace}
\newcommand{\kT}           {\ensuremath{k_{\rm T}}\xspace}
\newcommand{\px}           {\ensuremath{p_{\rm x}}\xspace}
\newcommand{\py}           {\ensuremath{p_{\rm y}}\xspace}
\newcommand{\etarange}[1]  {\mbox{$\left| \eta \right| < #1$}}
\newcommand{\yrange}[1]    {\mbox{$\left| y \right| < #1$}}
\newcommand{\dndy}         {\ensuremath{\mathrm{d}N_\mathrm{ch}/\mathrm{d}y}\xspace}
\newcommand{\dndeta}       {\ensuremath{\mathrm{d}N_\mathrm{ch}/\mathrm{d}\eta}\xspace}
\newcommand{\avdndeta}     {\ensuremath{\langle\dndeta\rangle}\xspace}
\newcommand{\dNdy}         {\ensuremath{\mathrm{d}N_\mathrm{ch}/\mathrm{d}y}\xspace}
\newcommand{\dEdx}         {\ensuremath{\textrm{d}E/\textrm{d}x}\xspace}
\newcommand{\kstar}        {\ensuremath{k^*}\xspace}
\newcommand{\mt}           {\ensuremath{m_{\rm{T}}}\xspace}
\newcommand{\St}           {\ensuremath{S_{\rm{T}}}\xspace}

\newcommand{\seven}        {$\sqrt{s}~=~7$~Te\kern-.1emV\xspace}
\newcommand{\thirteen}        {$\sqrt{s}~=~13$~Te\kern-.1emV\xspace}
\newcommand{\tev}          {Te\kern-.1emV\xspace}
\newcommand{\lumi}         {\ensuremath{\mathcal{L}}\xspace}

\newcommand{\ITS}          {\rm{ITS}\xspace}
\newcommand{\TOF}          {\rm{TOF}\xspace}
\newcommand{\ZDC}          {\rm{ZDC}\xspace}
\newcommand{\ZDCs}         {\rm{ZDCs}\xspace}
\newcommand{\ZNA}          {\rm{ZNA}\xspace}
\newcommand{\ZNC}          {\rm{ZNC}\xspace}
\newcommand{\SPD}          {\rm{SPD}\xspace}
\newcommand{\SDD}          {\rm{SDD}\xspace}
\newcommand{\SSD}          {\rm{SSD}\xspace}
\newcommand{\TPC}          {\rm{TPC}\xspace}
\newcommand{\TRD}          {\rm{TRD}\xspace}
\newcommand{\VZERO}        {\rm{V0}\xspace}
\newcommand{\VZEROA}       {\rm{V0A}\xspace}
\newcommand{\VZEROC}       {\rm{V0C}\xspace}
\newcommand{\Vdecay} 	   {\ensuremath{V^{0}}\xspace}

\newcommand{\ee}           {\ensuremath{\mathrm{e}^{+}\mathrm{e}^{-}}} 
\newcommand{\pip}          {\ensuremath{\pi^{+}}\xspace}
\newcommand{\pipi}          {\ensuremath{\pi^{\pm}\mbox{--}\pi^{\pm}}\xspace}
\newcommand{\pK}          {\ensuremath{\mathrm{p}\mbox{--}\mathrm{K}^{+}}\xspace}
\newcommand{\pim}          {\ensuremath{\pi^{-}}\xspace}
\newcommand{\kap}          {\ensuremath{\mathrm{K}^{+}}\xspace}
\newcommand{\kam}          {\ensuremath{\mathrm{K}^{-}}\xspace}
\newcommand{\KK}           {\ensuremath{\mathrm{K}^{+}\mathrm{K}^{-}}\xspace}
\newcommand{\KKbar}           {\ensuremath{\mathrm{K}\overline{\mathrm{K}}}\xspace}
\newcommand{\phiPart}      {\ensuremath{\phi}\xspace}
\newcommand{\pbar}         {\ensuremath{\overline{\mathrm{p}}}\xspace}

\newcommand{\dbar}{$\overline{\mathrm{d}}$}
\newcommand{\kzero}        {\ensuremath{{\mathrm K}^{0}_{\mathrm{S}}}\xspace}
\newcommand{\lmb}          {\ensuremath{\Lambda}\xspace}
\newcommand{\almb}         {\ensuremath{\overline{\Lambda}}\xspace}
\newcommand{\Om}           {\ensuremath{\Omega^-}\xspace}
\newcommand{\Mo}           {\ensuremath{\overline{\Omega}^+}\xspace}
\newcommand{\X}            {\ensuremath{\Xi^-}\xspace}
\newcommand{\Ix}           {\ensuremath{\overline{\Xi}^+}\xspace}
\newcommand{\Xis}          {\ensuremath{\Xi^{\pm}}\xspace}
\newcommand{\Oms}          {\ensuremath{\Omega^{\pm}}\xspace}

\newcommand{\Led}         {Lednick\'y--Lyuboshits\xspace}
\newcommand{\Ledn}         {Lednick\'y--Lyuboshits approach\xspace}
\newcommand{\Nphi}         {\ensuremath{\mathrm{N}}\mbox{--}\ensuremath{\phi}\xspace}
\newcommand{\pphi}         {\ensuremath{\mathrm{p}}\mbox{--}\ensuremath{\phi}\xspace}
\newcommand{\apphi}        {\ensuremath{\overline{\mathrm{p}}}\mbox{--}\ensuremath{\phi}\xspace}
\newcommand{\pphiComb}     {\ensuremath{\mathrm{p}}\mbox{--}\ensuremath{\phi} \ensuremath{\oplus} \ensuremath{\overline{\mathrm{p}}}\mbox{--}\ensuremath{\phi}\xspace}

\newcommand{\pd}{\ensuremath{\mbox{p--d}}\xspace}
\newcommand{\ApAd}         {\ensuremath{\mathrm{\overline{p}}\mbox{--}\mathrm{\overline{d}}}\xspace}
\newcommand{\ld}{\ensuremath{\mbox{$\Lambda$--d}}~}

\newcommand{\pKK}         {\ensuremath{\mathrm{p}}\mbox{--}(\ensuremath{\mathrm{K}^{+}\mathrm{K}^{-}})\xspace}

\newcommand{\pP}           {\ensuremath{\mathrm{p}\mbox{--}\mathrm{p}}\xspace}
\newcommand{\pL}           {\ensuremath{\mathrm{p}\mbox{--}\Lambda}\xspace}

\newcommand{\ApAP}         {\ensuremath{\mathrm{\overline{p}}\mbox{--}\mathrm{\overline{p}}}\xspace}

\newcommand{\pPComb}       {\ensuremath{\mathrm{p}\mbox{--}\mathrm{p} \oplus \mathrm{\overline{p}}\mbox{--}\mathrm{\overline{p}}}\xspace}

\newcommand{\pdComb}       {\ensuremath{\mathrm{p}\mbox{--}\mathrm{d} \oplus \mathrm{\overline{p}}\mbox{--}\mathrm{\overline{d}}}\xspace}

\newcommand{\pppair}           {\ensuremath{\mathrm{p}\mathrm{p}}\xspace}
\newcommand{\pdpair}           {\ensuremath{\mathrm{p}\mathrm{d}}\xspace}

\newcommand{\pppLair}           {\ensuremath{\mathrm{p}\mathrm{p}_{\Lambda}}\xspace}

\newcommand{\pLpair} {\ensuremath{\mathrm{p}\mbox{-}\Lambda}\xspace}

\newcommand{\Rpd} {\ensuremath{\mathrm{R}_\mathrm{pd}}\xspace}
\newcommand{\Rd} {\ensuremath{\mathrm{R}_\mathrm{d}}\xspace}
\newcommand{\Rp} {\ensuremath{\mathrm{R}_\mathrm{p}}\xspace}

\newcommand{\Singled} {\ensuremath{\mathrm{d}}\xspace}

\newcommand{\UGeVc}{GeV/$\it{c}$\xspace}
\newcommand{\UGeVcc}{GeV/$\it{c}^{2}$\xspace}
\newcommand{\nsigmaTPC}{$n_{\sigma,\text{TPC}}$\xspace}
\newcommand{\nsigmaTOF}{$n_{\sigma,\text{TOF}}$\xspace}
\newcommand{\Lednicky}{Lednick$\rm \acute{y}$\xspace}

\newcommand{\fReal} {\ensuremath{\Re(f_0) =\nolinebreak 0.85 \pm\nolinebreak 0.34\,(\mathrm{stat.}) \pm\nolinebreak 0.14\,(\mathrm{syst.})~\si{fm}}\xspace}
\newcommand{\fImag} {\ensuremath{\Im(f_0) =\nolinebreak 0.16 \pm\nolinebreak 0.10\,(\mathrm{stat.}) \pm\nolinebreak 0.09\,(\mathrm{syst.})~\si{fm}}\xspace}
\newcommand{\dZero} {\ensuremath{d_0 =\nolinebreak 7.85 \pm\nolinebreak 1.54\,(\mathrm{stat.})  \pm\nolinebreak 0.26\,(\mathrm{syst.})~\si{fm}}\xspace}

\newcommand{\NsigmaNoCorr} {\ensuremath{5.7 \pm\nolinebreak 0.8\,(\mathrm{stat.})  \pm\nolinebreak 0.5\,(\mathrm{syst.})~\sigma}\xspace }

\newcommand{\NsigmaNoCorrRange} {\ensuremath{4.7 - 6.6~\sigma}\xspace}

\newcommand{\VeffGauss} {\ensuremath{V_{\mathrm{eff}}=\nolinebreak 2.5\pm\nolinebreak 0.9 \,(\mathrm{stat.})\pm\nolinebreak 1.4\,(\mathrm{syst.})~\si{\MeV}}\xspace}
\newcommand{\muGauss}   {\ensuremath{\upmu=\nolinebreak0.14\pm\nolinebreak 0.06 \,(\mathrm{stat.})\pm\nolinebreak 0.09\,(\mathrm{syst.})~\si{fm^{-2}}}\xspace}

\newcommand{\alphaYuk}  {\ensuremath{\upalpha=\nolinebreak65.9 \pm\nolinebreak 38.0\,(\mathrm{stat.})\pm\nolinebreak 17.5\,(\mathrm{syst.})~\si{\MeV}}\xspace}
\newcommand{\AYuk}  {\ensuremath{A=\nolinebreak0.021\pm\nolinebreak 0.009\,(\mathrm{stat.})\pm\nolinebreak 0.006\,(\mathrm{syst.})}\xspace}
\newcommand{\gYuk}  {\ensuremath{g_{\Nphi} =\nolinebreak 0.14\pm\nolinebreak 0.03\,(\mathrm{stat.})\pm\nolinebreak 0.02\,(\mathrm{syst.})}\xspace}

\newcommand{\norm} {\ensuremath{\mathit{M} = 0.96}\xspace}

\newcommand{\realScatSigma}{\ensuremath{2.3\upsigma}\xspace}

\begin{titlepage}
\PHyear{2025}       
\PHnumber{096}      
\PHdate{22 April}  

\title{
Femtoscopic study of the proton-proton and proton-deuteron systems in heavy-ion collisions at the LHC}
\ShortTitle{Strong interaction studies in \pd system}   

\Collaboration{ALICE Collaboration\thanks{See Appendix~\ref{app:collab} for the list of collaboration members}}
\ShortAuthor{ALICE Collaboration} 

\begin{abstract}
This work reports femtoscopic correlations of $\pP\ (\ApAP)$ and $\pd\ (\ApAd)$ pairs measured in \mbox{Pb--Pb} collisions at center-of-mass energy per nucleon $\sqrt{s_{\rm NN}}$ = 5.02~TeV in the ALICE Collaboration.
A fit to the measured proton-proton correlation functions allows one to extract the dependence of the nucleon femtoscopic radius of the particle-emitting source on the pair transverse mass ($m_\text{T}$) and on the average charge particle multiplicity $\langle\text{dN}_\text{ch}/\text{d}\eta\rangle^{1/3}$ for three centrality intervals (0--10$\%$, \mbox{10--30$\%$}, 30--50$\%$). 
In both cases, the expected power-law and linear scalings are observed, respectively.  
The measured \pd correlations can be described by both two- and three-body calculations, indicating that the femtoscopy observable is not sensitive to the short-distance features of the dynamics of the p-(p-n) system, due to the large inter-particle distances in \mbox{Pb--Pb} collisions at the LHC.
Indeed, in this study, the minimum measured femtoscopic source sizes for protons and deuterons have a minimum value at $2.73^{+0.05}_{-0.05}$ and $3.10^{+1.04}_{-0.86}$~fm, respectively, for the 30--50$\%$ centrality collisions.
Moreover, the $m_{\rm{T}}$-scaling obtained for the \pP and \pd systems is compatible within 1$\sigma$ of the uncertainties.
These findings provide new input for fundamental studies on the production of light (anti)nuclei under extreme conditions.

\end{abstract}
\end{titlepage}

\setcounter{page}{2} 


\section{Introduction}
The energy density reached in ultra-relativistic heavy-ion collisions (HICs) allows for the production of short-lived droplets made of deconfined quarks and gluons~\cite{QGP1975,QGP1980,QGP2000,QGP_BRAHMS05,QGP_PHOBOS,QGP_STAR,QGP_PHENIX,ALICE:2022wpn}.
This state of matter, called the quark--gluon plasma (QGP), evolves in approximately $10 ~\mathrm{fm}/\it{c}$~\cite{10fmc_1999, 10fmc_2002} into final-state particles through the hadronization process, after which the quarks and gluons are confined inside hadrons~\cite{Letessier_Rafelski_2002}. 

In HICs, statistical hadronization models (SHMs)~\cite{SHMmodel, production1_276} can well reproduce the production yields of abundant hadron species such as $\pi$ and p, as well as light ions and particles containing heavier quarks, e.g., d, $ ^3\text{He}$, K, $\Lambda$, and $\phi$. 
These models describe the system that undergoes hadronization at a chemical potential close to zero and a temperature of approximately $156~\pm~9$ ~MeV~\cite{Tch156}.
This temperature value is also predicted by lattice quantum chromodynamics calculations~\cite{LQCD} for the cross-over of the QGP system into a hadronic system.
It is surprising that the models also work well in predicting the yield of nucleon-based composite objects, such as deuterons and other light nuclei or even hypernuclei, despite the large difference between the temperatures that characterize the hadronization process and the binding energy of these particles. 
For example, the binding energy of the deuteron is 2.22~MeV~\cite{d_bindE}, 70 times smaller than the chemical equilibrium temperature extracted employing the SHMs. 
Alternative descriptions of data on nuclei are based on the coalescence mechanism~\cite{CoalInHIC,kaijia_dHe3}. 
It describes the formation of light nuclei in high-energy hadron collisions by combining nucleons that are close in phase space during the expansion of the fireball. 
Calculations based on the coalescence
mechanism have shown significant success in reproducing experimental data on light-nucleus production from measurements conducted at the Large Hadron Collider (LHC) and the Relativistic Heavy Ion Collider (RHIC), demonstrating its effectiveness in capturing the dynamics of nuclear formation in HICs~\cite{pro3,production7}. 

In addition to the aforementioned spectra studies, light ion production mechanisms can be investigated through the measurement of elliptic flow, which is sensitive to the conditions in the early stages of the system evolution~\cite{flow6, PhysRevC.94.034908, production7, production77}. 
However, the elliptic flow distribution of deuterons could not be consistently reproduced across all centralities using only SHMs or the coalescence approach~\cite{production77}. 
Additionally, the production of deuterons in HICs should be reflected in fluctuation measurements~\cite{production4444}, the results of which have been well reproduced by SHMs but not by simple coalescence models. 
Despite the combined efforts across different observables, the overall understanding of deuteron production in collisions involving the QGP state remains ambiguous.

Furthermore, momentum correlations involving nuclei can be exploited to study the production mechanisms of light ions~\cite{Kachelriess:2019taq,Blum:2019suo,Mahlein:2023fmx,Mrowczynski2019yrr} by comparing their emission properties with those of abundant hadrons emitted in the collisions. 
Femtoscopy, which relies on analyzing momentum correlations between particles, is used to study the space-time properties of particle emission and to examine the final-state interactions (FSI) of the emitted hadrons~\cite{Lisa:2005dd}.
Measurements carried out in HICs for particles which are formed directly in the hadronization process, such as $\pi$, K, and p~\cite{ALICE:2011kmy,ALICE:2010igk,ALICE:2015hav,ALICE:2015tra,ALICE:2020mkb,alice276}, showed a linear dependence of the femtoscopic radii with the cube root of the mean charged-particle multiplicity~\cite{STAR2001,PHENIX2004,star2009,Alice2011,alice276,ppSource} and a decreasing power-law dependence upon the pair transverse mass, $\mt$~\cite{STAR2005,alice276}. 
For single particles, the latter is defined as $\mt = \sqrt{p_{\rm T}^{2} + m^{2}}$, where $p_{\rm T}$ is the particle’s transverse momentum and $m$ is its rest mass. 
For identical particle pairs, $p_{\rm T}$ in this expression is replaced by $\kT = \left|\boldsymbol{p}_{{\mathrm{T}, 1}} + \boldsymbol{p}_{{\mathrm{T}, 2}}\right| / 2$, which is the average transverse momentum of the pair.
The observed \mt-scaling is attributed to the radial flow of the collective expansion of the system driven by internal pressure.
Indeed, these effects can be replicated by models that include a QGP phase described through relativistic hydrodynamic calculations~\cite{therm2014,therm2018, yu2019}. 
A direct consequence of the hydrodynamic evolution of the system is the observation that the higher the transverse mass of the particle, the smaller the femtoscopic source size~\cite{PhysRevC.81.064906}.
The common \mt-scaling has been observed for mesons and baryons such as $\pi$, K, and p both in pp and \pbpb collisions~\cite{ppSource,SourceMaxi,alice276,Marcel_ppi,ppSource}. However, it remains an open question, whether light nuclei and hypernuclei follow the common \mt-scaling of the other hadrons.

This paper presents measurements of the space-time properties of the emission of nucleons and deuterons as a function of pair \mt based on the femtoscopy of two identical protons (\pP) and femtoscopy of proton--deuteron pairs (\pd).
In both \pP and \pd analyses, three centrality intervals were chosen \mbox{0--10$\%$}, 10--30$\%$, and 30--50$\%$. 
The most precise measurements of \pP correlation functions in eight \mt ranges in \pbpb collisions at $\sqrt{s_{\rm NN}}=5.02$~TeV are presented, aimed at determining the \mt scaling of nucleon femtoscopic radii.
The \pd correlations were then interpreted using two different models for the FSI~\cite{Lednicky:1981su,Viviani2023} to extract the corresponding single nucleon and deuteron source radii in two distinct \mt ranges. 
This analysis implies that deuterons follow the same \mt-scaling behavior observed for single nucleons.

\section{Data analysis}
\label{sec:selection}
The study is based on $\num{337e6}$  Pb--Pb collisions at $\sqrt{s_{\rm NN}}=5.02$~TeV collected during the LHC Run 2 period (2018) with the ALICE detector~\cite{ALICE:2008ngc, ALICE:2014sbx}. 
The main sub-detectors used in this work are the V0 detectors and three central barrel detectors: the Inner Tracking System (ITS)~\cite{Aamodt:2010aa}, the Time Projection Chamber (TPC)~\cite{Gasik:2018ush}, and the Time-Of-Flight detector (TOF)~\cite{Gasik:2018ush}. 
The V0 detectors (V0A and V0C)~\cite{Abbas:2013taa} are located outside the central barrel on both sides of the interaction point.
All the central barrel detectors are placed in a homogenous $0.5$~T solenoidal magnetic field that is parallel to the beam direction.

The V0 detectors are utilized in the trigger system to determine the centrality of the collision. 
They consist of two arrays of scintillators, each comprising 32 scintillator counters, covering the pseudorapidity ranges of $2.8 < \eta < 5.1$ (V0A), and $-3.7 < \eta < -1.7$ (V0C). 
The minimum bias trigger for recording collision events requires that both V0 detectors register at least one hit simultaneously, coinciding with a collision occurring within a bunch crossing at the center of the ALICE detector. 
The centrality of the collision, expressed as a percentage of the total hadronic cross section, is determined based on the amplitudes of the signals left in both V0 arrays~\cite{Abelev:2013qoq}. 
In this study, the collisions are classified into three intervals of the Pb–Pb cross section: $\mbox{0--10\%}$, $\mbox{10--30\%}$, and $\mbox{30--50\%}$. All centrality classes use events selected using the minimum bias trigger. 
In addition, the central and semi-central triggers~\cite{ALICEcentrality} are used to enhance the statistical significance in $\mbox{0--10\%}$ and $\mbox{30--50\%}$ centrality intervals, respectively. 
To ensure uniform acceptance at midrapidity, only collisions occurring within $\pm~10$ cm of the nominal interaction point along the beam axis are used in the study. 

The ITS is a silicon detector made of six cylindrical layers used for precise primary vertex reconstruction. 
It also participates in pile-up removal by utilizing hit information from the SPD layers, resulting in a negligible residual pile-up contribution~\cite{Aamodt:2010aa, ALICE:2014sbx}.
The TPC is the main tracking detector, used for momentum reconstruction and particle identification (PID) via specific energy-loss (d$E$/d$x$) measurement. 
It is a gas-filled cylindrical barrel, divided into two halves by a central cathode, with each half having its readout. 
Charged particles ionize the gas, producing drifting electrons that create signals in the pad rows of one of 18 azimuthal sectors. 
The reconstructed tracks are required to have at least 80 clusters deposited in the TPC by a single particle.
The TOF detector is also used in PID, as it can determine the mass hypothesis of a measured particle by combining its time-of-flight measurement with the expected trajectory length and momentum. 
It is shaped as a hollow cylinder surrounding the other detectors at a distance of about 380 cm from the center of the beam pipe. 
The TOF utilizes Multigap Resistive Plate Chamber (MRPC) detectors to measure the time at which charged particles cross its sensitive gas volume.

The PID is based on evaluating absolute deviations from the signal hypotheses, denoted $n_{\sigma,\text{TPC}}$ and $n_{\sigma,\text{TOF}}$, expressed in units of the respective detector resolutions at specific particle momentum. 
For this analysis, different transverse momentum intervals were considered for the tracks, depending on the particle species. The $p_{\rm T}$ ranges are $\mbox{0.5--3.0}$~\UGeVc for (anti)protons and $\mbox{0.8--2.0}$~\UGeVc for (anti)deuterons. 
The (anti)proton candidates are selected by combining TPC and TOF signals, requiring \mbox{$\sqrt{n_{\sigma,\text{TPC}}^2+n_{\sigma,\text{TOF}}^2}<3$}. To ensure high-quality identification of (anti)deuterons, the strict criterion of $n_{\sigma,\text{TPC}}< 2$ is applied for momenta $p<1.3$ \UGeVc, while above this range, both $n_{\sigma,\text{TPC}} < 2$ and $n_{\sigma,\text{TOF}} < 2$ are employed simultaneously. 
The particles under study are identified in the pseudorapidity range $|\eta|<0.8$. 
To reduce the contribution of (anti)protons and deuterons candidates originating from weak decays or knock-out occurring in the interaction of particles with the detector material, the distance of the closest approach (DCA) to the primary vertex is required to be smaller than $0.0105 +  0.0350/(p_{\text{T}}/(\rm{GeV}/\it{c}))^\mathrm{1.1} ~\text{cm}$ in the transverse plane and 1 cm along the beam direction~\cite{kpinPbPb}, as most secondary particles have larger DCAs. 
The contribution of (anti)deuterons originating from weak decays is negligible, and there are no secondary antideuterons from the material. Therefore, antideuteron tracks are accepted if the DCA is smaller than 2.4~cm in the transverse plane and 3.2~cm along the beam direction, which corresponds to loose spatial criteria that maximize the efficiency for detecting primary particles originating from the event vertex.

Finally, the selected (anti)proton and (anti)deuteron candidates are combined to build proton--proton, antiproton--antiproton (\ApAP), proton--deuteron, and antiproton--antideuteron (\ApAd) pairs. 
To ensure good track quality and suppress the effect of possible multiple (cloned or split) tracks originating from the signal left by a single particle, reconstructed track pairs are required to share less than 5$\%$ of their clusters.
Furthermore, to minimize detector effects, such as track merging, a similar approach based on a close-pair-rejection criterion~\cite{kpinPbPb} is applied when the relative pseudorapidity of the two tracks is less than 0.01 ($|\Delta \eta|<0.01$). 
Pairs of tracks within the volume of the TPC are removed if their spatial overlap, defined as the percentage of points where the distance between two tracks in TPC is smaller than 3~cm, is above 5$\%$ and 2$\%$ for \pP and \pd study, respectively.
The pair selection criteria are optimized individually for each analysis to enhance the purity and quality of the selected pairs, while maintaining sufficient statistical precision.

\section{The femtoscopic correlation function}
\label{sec:cf_formula}
The femtoscopic correlation function, denoted as $C(\kstar)$, is measured as a function of the relative momentum. 
Here, $\kstar$ is defined as $\kstar = |\boldsymbol{p}^*_1-\boldsymbol{p}^*_2|/2$, where $\boldsymbol{p}^*_{1/2}$ are the particle momenta in the pair rest frame, indicated by the asterisk ($*$) notation. Experimentally, it is computed as
\begin{equation}
    C_\text{exp}(\kstar)=\mathcal{N}\frac{N_{\rm{same}}(\kstar)}{N_{\rm{mixed}}(\kstar)},
\end{equation}
where $N_{\rm{same}}(\kstar)$ is the correlated distribution of \kstar, which is obtained from particle pairs originating in the same collision (same event), and $N_{\rm{mixed}}(\kstar)$ is the corresponding distribution of uncorrelated pairs. 
The latter is generated using the mixed-event technique~\cite{Lisa:2005dd}. Mixed pairs are constructed by selecting particles from different events with similar primary vertex positions -- difference less than 2~cm, and similar centrality classes -- specifically, a centrality difference of less than 2.5$\%$ for central collisions (0--10$\%$) and less than 5$\%$ for semi-central collisions (10--30$\%$ and 30--50$\%$). Each event is mixed with 10 others.
The normalization term~$\mathcal{N}$ is used to scale the mixed-event distribution such that the integral over the selected range is equivalent to the analogous integral calculated for the same events. 
The~$\mathcal{N}$ factor is obtained within the range $\kstar\in(300,600)$ MeV$/c$ for \pP and \pd pairs. 
This interval is selected to ensure the absence of a femtoscopic signal, i.e. when $C(\kstar)$ approaches unity. The number of pairs used in the analysis that contribute to the $\kstar <$ 200~MeV/$c$ interval, is summarized in Tab.~\ref{tab:numbers}.

\begin{table}[h!]
 \caption{Number of \pP, \ApAP, \pd, and \ApAd pairs originating in the same collisions and contributing to the $\kstar <$ 200~MeV/$c$ interval.}
\centering
{ 
\begin{tabular}{ c | c | c | c | c }
 \hline
  Centrality & \pP  & \ApAP  & \pd  & \ApAd  \\ \hline \hline
0-10\%	& $\num{1.5e8}$ & $\num{9.0 e7}$ & $\num{6.0e4}$  & $\num{2.8e4}$ \\ \hline
10-30\%	& $\num{1.5e7}$ & $\num{9.6e6}$ & $\num{7.0e3}$ & $\num{3.5e3}$ \\ \hline
30-50\%	& $\num{1.2e7}$ & $\num{7.2e6}$ & $\num{5.6e3}$ & $\num{3.0e3}$\\ \hline

\end{tabular}
}
  \label{tab:numbers}
\end{table}

The relation between the experimental $C_\text{exp}(\kstar)$ and model $C_\text{femto}(\kstar)$ correlation function can be expressed as
\begin{equation}
C_\text{exp}(\kstar)= C_\text{baseline}(k^{*})\cdot C_\text{femto}(k^{*}),
\label{eq:expfemtobase}
\end{equation}
where $C_\text{baseline}(\kstar)$ describes the non-femtoscopic background which accounts for any residual contributions in $C_{\text{exp}}(\kstar)$. 
In this study, the background in the femtoscopic region primarily arises from elliptic flow~\cite{Kisiel:2017gip}, which is a consequence of the event-wise angular distribution of particles. While deuteron production may be enhanced within jets due to coalescence, the dominant source of deuterons in central and semi-central Pb–Pb collisions is the underlying event. Most particles in such collisions are emitted from a large, thermalized source exhibiting collective behavior, rather than from independent jet fragmentation. Given the strong jet quenching in these collisions, any minijet-related contributions to the femtoscopic correlation functions are expected to be negligible.
The handling of the non-femtoscopic background in the analysis depends on the fitting method and is described in the corresponding sections of the \pP and \pd studies in the text below.

The model correlation function accounts not only for the genuine correlations originating from pairs produced in primary processes after collisions, but also for the distortion of the measured correlation function due to particle misidentification and for secondary particles resulting from weak decays and spallation processes in the detector material.
These contributions are taken into account by calculating the model correlation function $C_\text{femto}(k^{*})$ as
 \begin{equation}
C_{\text{femto}}(\kstar)=1+\lambda_{\text{gen}}\left(C_{\text{gen}}(\kstar)-1\right)
+\lambda_{\text{feed}}\left(C_{\text{feed}}(\kstar)-1\right)
+\lambda_{\text{misid}}\left(C_{\text{misid}}(\kstar)-1\right),
\label{eq:Cfemto}
\end{equation}
which takes into account all contributions with the method discussed in Ref.~\cite{ppSource}. In particular, the contributions of genuine and non-genuine particles are quantified by the so-called $\lambda$ parameters, which can be evaluated as
\begin{equation}
\label{eq:wzor}\lambda_{\text{}}=\mathcal{P}_{i}\mathcal{F}_{i}\mathcal{P}_{j}\mathcal{F}_{j}.
\end{equation} 
In this formula, $\mathcal{P}$ represents the PID purity of a given particle species accepted for the study, and $\mathcal{F}$ denotes the fraction of particles of interest with either primary or secondary origin. 
Both quantities are obtained separately for the two particles contributing to the pair, denoted by the subscripts $i,j$. 
The fractions are calculated using approaches already employed in other femtoscopy studies~\cite{kpinPbPb, Bhawanipd}. 
The identification purity of the (anti)proton is estimated from Monte Carlo (MC) simulations using the HIJING~\cite{Wang:1991hta} event generator coupled with the GEANT3~\cite{Brun:1994aa} transport code. 
The fraction of correctly identified and misidentified particles within an accepted sample of (anti)deuterons is calculated using a data-driven approach that incorporates information on signals in the TPC and TOF detectors. 
To estimate the primary and secondary contributions to the (anti)proton samples, a combined MC and data-driven method involving template fits to the $\rm{DCA}_{\it{xy}}$ distributions is employed. 
The primary fraction of antideuterons, $\mathcal{F}$, is assumed to be 100\%, as such particles cannot be produced via material knock-out due to baryon number conservation. 
The average $\mathcal{P}$ and $\mathcal{F}$ fractions for (anti)protons and (anti)deuterons are listed in Tab.~\ref{tab:PurityAndFrac}. Consequently, the genuine component fraction of $C_\text{femto}(k^{*})$ for correctly identified primary particle pairs corresponds to $\lambda_{\text{gen}}$, the residual contribution from feed-down of weak decays corresponds to $\lambda_{\text{feed}}$, and the residual contributions of misidentified particles is given by $\lambda_{\text{misid}}$. 
The functional form of $C_{\text{gen}}$ is obtained from the model description of the interaction of the pair under study, $C_{\text{misid}}$ is assumed to be flat and $C_{\text{feed}}$ is obtained by projecting the contribution of weak decays onto the $k^*$ of the genuine pairs. 
More details about the $\lambda$ methodology can be found in Refs.~\cite{SourceMaxi, PhysRevC.99.024001}. 
To better reflect the properties of the data samples, the genuine $\lambda_\text{gen}$ parameters of \pP and \pd pairs are computed as a function of $m_\text{T}$ and $k^*$, respectively. 
After applying the $\lambda$ corrections, no systematic difference is observed between particle and antiparticle pairs within the uncertainties. 
Since no difference is expected in their interactions, the distributions of particle and antiparticle pairs are combined to reduce the statistical uncertainty.
Therefore, in the following text, the terms \pP and \pd refer to the combination of $\pP \oplus \ApAP$ and  $\pd\oplus\ApAd$, respectively.
\begin{table}[t!]
    \centering
    \caption{The averages of the PID purity $\mathcal{P}_{\text{PID}}$, the primary fraction $\mathcal{F}_\text{primary}$, feed-down contribution of weak decays $\mathcal{F}_\text{feed-down}$, and material budget $\mathcal{F}_\text{material}$ for (anti)protons and (anti)deuterons. The values are presented as left- and right-range values, corresponding to the $0\mbox{--}10\%$ and $30\mbox{--}50\%$ centrality intervals, respectively.}     
    
    \begin{tabular}{c|c|c|c}
        \hline
         & $\rm{p}(\rm{\bar{p}})$ & $\rm{d}$ & $\rm{\bar{d}}$\\
         \hline\hline
        $\mathcal{P}_{\text{PID}}$ & $97\mbox{--}99\%$ & $93\mbox{--}98\%$ & $93\mbox{--}98\%$\\
        \hline
        $\mathcal{F}_\text{primary}$ & $88\mbox{--}90\%$ & $74\mbox{--}89\%$ & $100\%$\\
        \hline   
        $\mathcal{F}_\text{feed-down}$ & $ 12\mbox{--}10\%$ &  $-$  & $-$ \\
        \hline  
        $\mathcal{F}_\text{material}$ & $<1\%$ & $26\mbox{--}11\%$ & $-$\\
        \hline           
    \end{tabular}
    
    \label{tab:PurityAndFrac}
\end{table}

From the theoretical perspective, the two-particle correlation function is defined via the Koonin--Pratt formula~\cite{Lisa:2005dd, Fabbietti:2020bfg}:

 \begin{equation}
     C(k^*) = \int S(\boldsymbol{r}^*) |\psi(\boldsymbol{r}^{*},\boldsymbol{k}^*)|^2 \text{d}^3 r^* \,, \label{Eq. Koonin-Pratt}
 \end{equation}
where $\boldsymbol{r}^*$ describes the relative distance between the two particles in the pair rest frame and $\psi(\boldsymbol{r}^{*},\boldsymbol{k}^*)$ is the two-particle wave function.
The source function $S(\boldsymbol{r}^{*})$ is related to the probability of producing two particles at a relative distance $r^*$, and it can be modeled using a Gaussian profile:

\begin{equation}
S(\boldsymbol{r}^{*})=\frac{1}{\left(2 \pi R_{ij}^2\right)^{3 / 2}}  \exp{\left(-\frac{r^{* 2}}{2R_{ij}^2}\right)}.
\end{equation}

Here, the source size, referred to as radius, is characterized by the two-particle Gaussian source width of the distribution of the $i\mbox{--}j$ pair, $R_{ij}$, where $i$ and $j$ indicate the particle species forming the pair (e.g., $pp$ for proton–proton and $pd$ for proton–deuteron pairs).
In the case of identical particles, \mbox{$R_{ii} = \sqrt{2} R_{i}$} where $R_{i}$ denotes the single-particle femtoscopic source radius.
In pp collisions, due to the small source size ($R_{i}\sim$1.5~fm), the contributions of short-lived (strongly decaying) resonances ($\it{c}\tau\leq$ 5 fm) effectively increase the femtoscopic source size by introducing exponential tails to the source function. 
This effect is known as "resonance halo"~\cite{ppSource,SourceMaxi}. 
This effect can be neglected for sources of $R_{i}\sim3\mbox{--}8$ fm, typical for \pbpb collisions, as the convolution of the exponential with a Gaussian does not distort the width of the distribution within the sensitivity of the measurement. 
Thus, this analysis extracts the effective femtoscopic source radii.

\subsection{The proton--proton correlation function and nucleon \texorpdfstring{\mt-scaling}{mt-scaling}}
The experimental \pP correlations are investigated and fitted in the $k^*$ range up to 300~MeV/$c$, across eight different \mt intervals ($1.0\mbox{--}1.3$, $1.3\mbox{--}1.4$, $1.4\mbox{--}1.5$, $1.5\mbox{--}1.6$, $1.6\mbox{--}1.7$, $1.7\mbox{--}1.8$, $1.8\mbox{--}2.0$, $2.0\mbox{--}3.2$ GeV/$\it{c}^{2}$) and three centrality classes using Eqs.~\eqref{eq:expfemtobase} and~\eqref{eq:Cfemto}. 
The genuine part of the \pP correlation function, with the contribution $\lambda_{\text{gen}}$ calculated as described for three multiplicity classes mentioned in Sec.~\ref{sec:cf_formula}, is computed employing the CATS~\cite{Dimi_CATS} framework that includes the Argonne $\it{v}_\text{18}$ (AV18) potential~\cite{Av18}, the Coulomb interaction, and the Pauli-blocking effect between the two identical fermions. 
The $C_\text{baseline} (k^{*})$ describing the non-femtoscopic background of \pP pairs is modeled using a linear function as a part of the correlation function fit.

To model the experimental \pP correlation function, residual correlations from the feed-down of \pL pairs are explicitly considered using Chiral Effective Field Theory at next-to-leading order ($\chi$EFT at NLO)~\cite{NLNSigma,HN_NLO13,HN_NLO19}. 
The effect of the \pL correlation function on the \pP correlation is included by projecting~\cite{ppSource} this contribution onto the \kstar of \pP pairs via a decay matrix~\cite{pp_pL_LLinpp7TeV,pLinSTAR}, which is built according to the kinematics of the decay. 
The contribution from the feed-down of \pL pairs to \pP pairs is at most 2\%. All other feed-down and misidentification contributions are considered to be flat. 
All contributions in $C_\text{femto}(k^{*})$ are smeared to account for the finite momentum resolution of the ALICE detector~\cite{pp_pL_LLinpp7TeV}.

\begin{figure}[t!]
\centering
\includegraphics[width=0.95\textwidth]{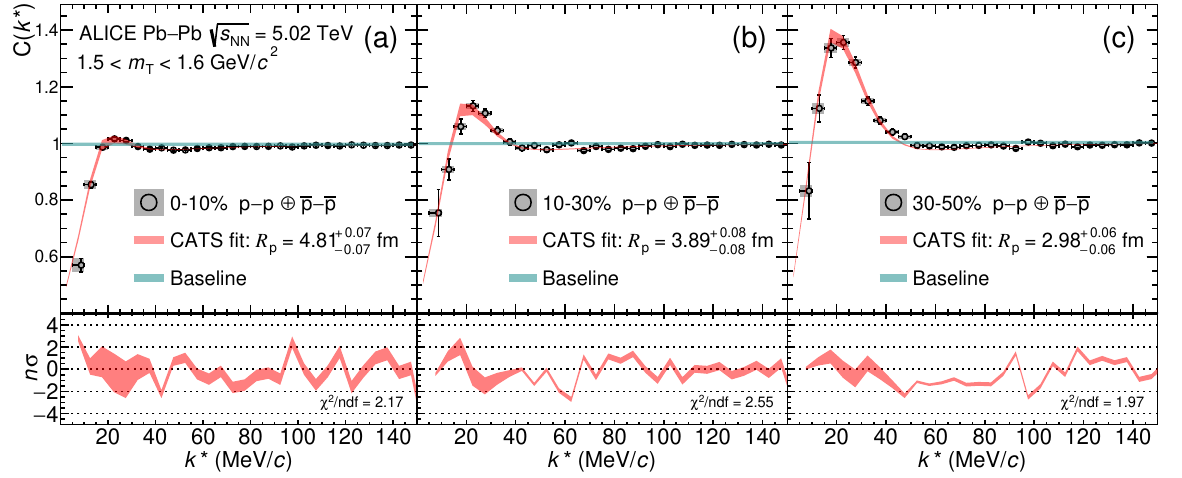}
\caption{Top panels: raw \pP correlation functions as a function of $k^{*}$ in one exemplary \mt interval obtained from Pb--Pb collisions at $\sqrt{s_{\rm NN}} = $ 5.02~TeV in the 0--10$\%$ (a), 10--30$\%$ (b) and 30--50$\%$ (c) centrality intervals, respectively. 
The black open markers represent the data, while the vertical lines and the boxes indicate the statistical and systematic uncertainties, respectively. 
The red region represents the 1$\sigma$ uncertainty range of the fits performed using CATS~\cite{Dimi_CATS}.
The non-femtoscopic background contributions are shown by the light cyan bands. 
Bottom panels: $n_{\sigma}$ calculated as data-to-fit difference, normalized by the total uncertainty of experimental data. 
}

\label{fig:ppCorr}
\end{figure}

The measured \pP correlation function employed to extract the \mt-scaling of $R_\text{p}$ is shown in Fig.~\ref{fig:ppCorr} for one representative \mt interval. The vertical bars and grey boxes represent the statistical and systematic uncertainties of the data, respectively.
The systematic uncertainties of the \pP correlation function data points are calculated by varying event and track selection criteria.
This includes changing the distance of the primary vertex to the center of the ALICE detector in the beam direction from $\pm~10$~cm to $\pm~7$~cm, the PID selection from 3$\sigma$ to 2.5$\sigma$ or 3.5$\sigma$, $|\eta|$ range from 0.8 to 0.77 and 0.83, the minimal $p_\text{T}$ range from 0.7~GeV/$c$ to 0.6 or 0.8~GeV/$c$, and the number of TPC space points required in the reconstruction process of the tracks from 80 to 70 or 90. The final uncertainty is estimated as the difference between the maximum and minimum values of the correlation function divided by the standard deviation of the uniform distribution.
The fitted $C_\text{exp}(\kstar)$ functions are shown in Fig.~\ref{fig:ppCorr} by the red bands and the obtained baseline contributions $C_\text{baseline}(\kstar)$ are shown as light cyan bands. The resulting baseline shows no significant contribution to the correlation function in all considered $m_{\rm T}$ bins.
The transition from central to peripheral collisions in a given \mt interval leads to a smaller emission source size, which visibly impacts the strength of the correlation function. 
This behavior is accurately captured by the fitting procedure currently employed.

\begin{figure}[t!]
    \centering
    \begin{minipage}{0.49\textwidth}
        \centering
    \includegraphics[width=\textwidth]{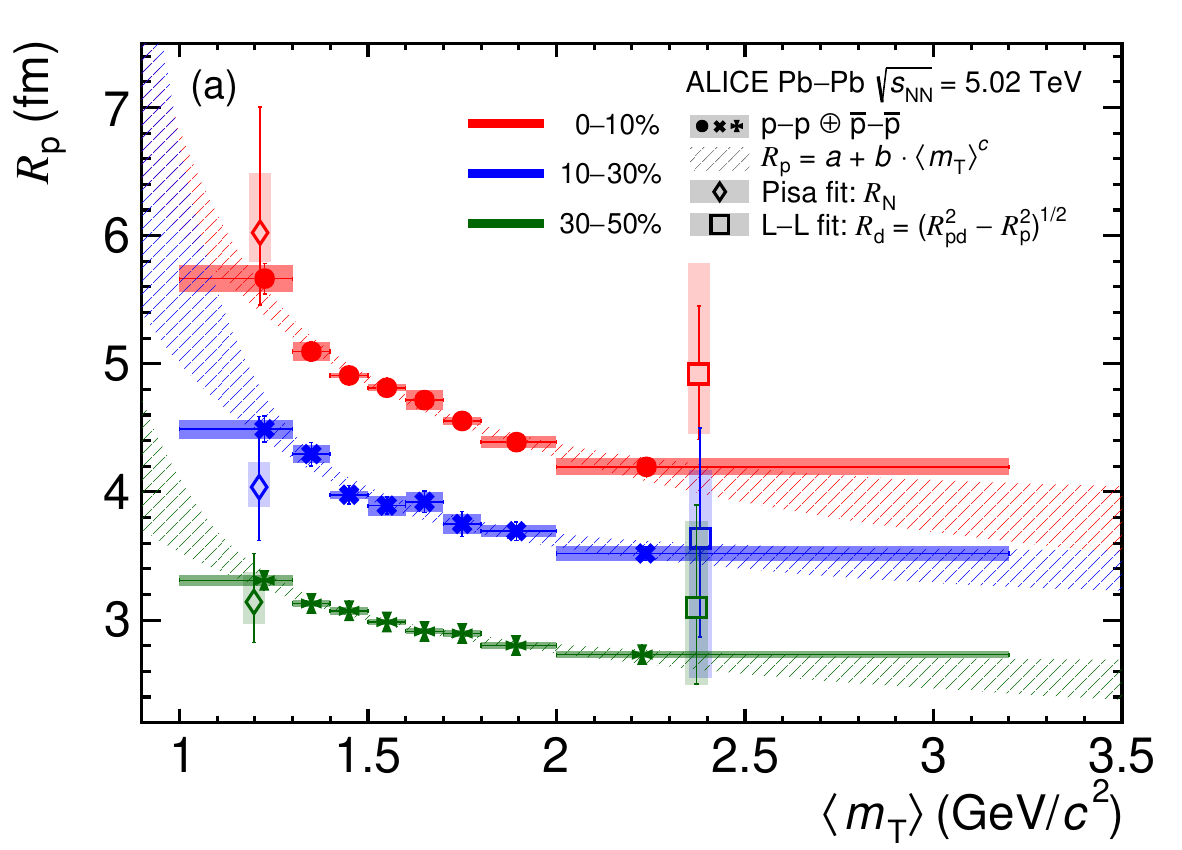}
        
    \end{minipage}%
    \hfill 
    \begin{minipage}{0.49\textwidth}
        \centering
    \includegraphics[width=\textwidth]{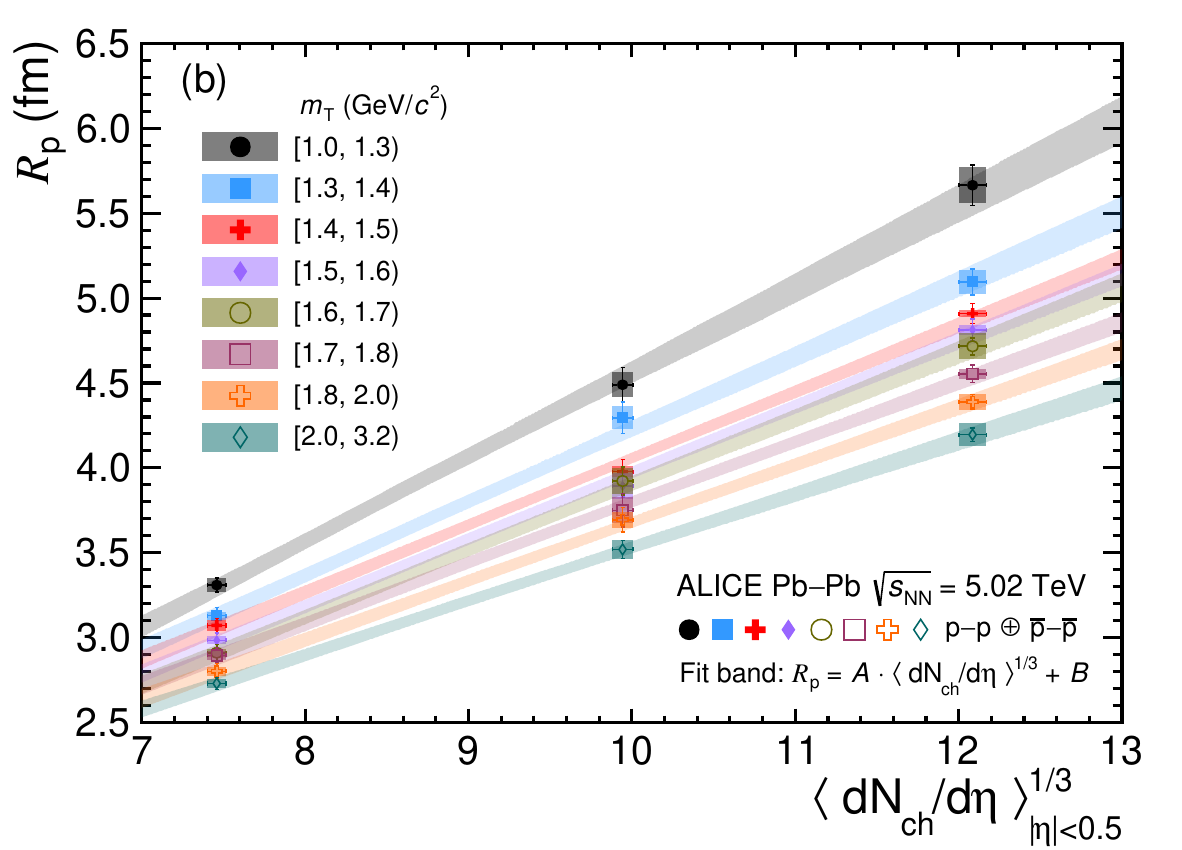}
        
    \end{minipage}
    \caption{Left panel: The femtoscopic source size of protons ($R_\text{p}$) measured as a function of $m_\text{T}$ in Pb--Pb collisions at $\sqrt{s_{\rm NN}}~=~$5.02~TeV is represented by full red, blue, and green markers for the \mbox{0--10$\%$}, \mbox{10--30$\%$}, \mbox{30--50$\%$} centrality intervals, respectively. 
    The open diamond and squared markers represent the values of femtoscopic source sizes of nucleons ($R_\text{N}$) and deuterons ($R_\text{d}$),  obtained by fitting the measured \pd correlation function with the L--L~\cite{Lednicky:1981su} and Pisa~\cite{Viviani2023} models, where in the later the AV18~\cite{Av18}+UIX~\cite{Urbana9} potential is used.
    Right panel: $R_\text{p}$ as a function of $\langle \text{dN}_\text{ch}/\text{d}\eta\rangle^{1/3}$ for eight \mt intervals. 
    The shaded bands in the two panels depict the $m_{\rm T}$ and $\langle \text{dN}_\text{ch}/\text{d}\eta\rangle^{1/3}$ dependence of the measured radii fitted with power-law function, $R_\text{p} = \it{a} + \it{b}\cdot\langle m_\text{T}\rangle^{\it{c}}$, and linear function $R_\text{p}=A\cdot\langle \text{dN}_\text{ch}/\text{d}\eta\rangle^{1/3}+B$, respectively. The bandwidth reflects the 1$\sigma$ range of the fit solutions, obtained by bootstrapping the radii over their statistical and systematic uncertainties. }
    
    \label{fig:MoneyPlot}
\end{figure}

The $R_\text{p}$ values obtained from fitting the \pP correlation function across each \mt interval are displayed in the left panel of Fig.~\ref{fig:MoneyPlot}. 
These values are shown for the 0--10$\%$, 10--30$\%$, and 30--50$\%$ centrality intervals, represented by solid red, blue, and green markers, respectively. 
The measured femtoscopic source sizes are relatively large, ranging from a minimum of $2.73^{+0.05}_{-0.05}$~fm for the highest \mt bin in the 30–50$\%$ centrality collisions to $5.66^{+0.16}_{-0.16}$~fm for the lowest \mt bin in the 0–10$\%$ centrality collisions. 
The values are quoted with combined systematic and statistical uncertainties.
The systematic uncertainties of all $R_\text{p}$
are estimated as 1$\sigma$ of the source radii distribution calculated for different settings of the analysis and fitting method. 
The uncertainties contributing from the experimental correlation functions are accounted by the bootstrap method~\cite{bootstrap} where data points are resampled using a Gaussian distribution of the statistical and systematic error. 
The uncertainty originating from the fitting technique is accounted for by varying the predefined fitting settings. To evaluate the possible impact of the fit range on the stability of the constrained baseline, the upper limit of the fit range was modified by $\pm~$50 MeV/c. 
Additionally, the baseline was modeled using either a first- or third-degree polynomial with a fixed linear term to accommodate potentially more complex shapes of non-femtoscopic effects. 
Finally, the $\lambda_{\rm gen}$ parameters are varied by up to $\pm~5\%$ to address potential underestimation or overestimation in the calculation of the genuine particle fractions in the sample. 
This $\pm~5\%$ variation is a conservative estimate intended to cover the full range of uncertainty arising from changes in particle purities and primary fractions due to variations in selection criteria used in systematic checks, as well as from uncertainties in the template fits to DCA distributions and the purity estimations obtained via data-driven methods.
Shaded bands in the left panel represent the 1$\sigma$ uncertainties of the femtoscopic source sizes dependence of $\mt$, described using a power-law parametrization,
$R_{\mathrm{p}} = \mathit{a} + \mathit{b}\cdot\langle m_{\mathrm{T}}\rangle^{\mathit{c}}$. The obtained values of the parameters $a,b,c$ are summarized in Appendix~\ref{sec:appendix:param}. 
The values of $R_{\text{p}}$ exhibit the same trend with increasing $m_{\text{T}}$ observed in previous femtoscopy analyses of $\pi$, K, and p in \pbpb collisions at $\sqrt{s_{\rm NN}}= 2.76$ TeV~\cite{alice276}. 
The $m_{\text{T}}$ scaling presented in the figure also includes nucleon-systems measurements from a proton-deuteron study, based on three-body (denoted in the figure as PISA) and two-body (Lednick\'y--Lyuboshitz approach denoted as \mbox{L--L}) analyses, which are described in more detail in the next section. 
Furthermore, the measured $R_{\text{p}}$ grows linearly with the cube root of the charged particle multiplicity for all \mt ranges separately, as shown in the right panel of Fig.~\ref{fig:MoneyPlot}. 
The bands of this panel depict 1$\sigma$ uncertainties of fits to the $\langle \text{dN}_\text{ch}/\text{d}\eta\rangle^{1/3}$ dependence of $R_\text{p}$, obtained using $R_{\text{p}}=\textit{A}\cdot\langle \text{dN}_\text{ch}/\text{d}\eta\rangle^{1/3}+\textit{B}$. The obtained values of the parameters $A, B$ are summarized in Appendix~\ref{sec:appendix:param}.

\subsection{The proton--deuteron correlation function }

\begin{figure*}[t!]
\centering
\includegraphics[width=0.95\textwidth]{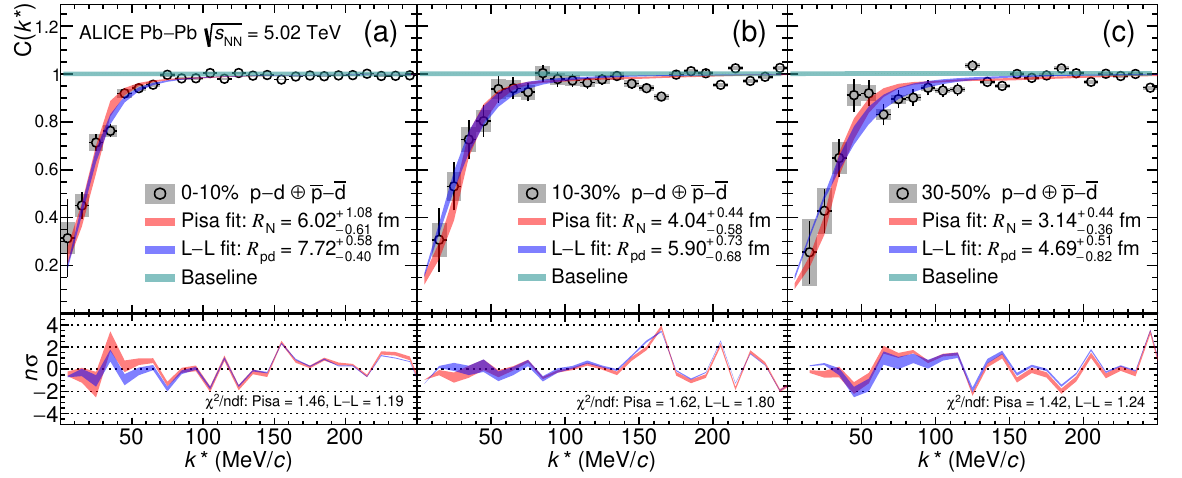}

\caption{Top panels: genuine \pd correlation function as a function of $k^{*}$ obtained from Pb--Pb collisions at $\sqrt{s_{\rm NN}} = $ 5.02~TeV in the 0--10$\%$ (a), 10--30$\%$ (b) and 30--50$\%$ (c) centrality intervals, respectively. The measured data are shown as black open markers, while the vertical lines and the boxes represent the statistical and systematic uncertainties, respectively. The correlation functions are shown together with a fit band of 1$\sigma$ uncertainties performed using the three-body approach employing AV18+UIX potentials (Pisa fit, red bands) and using the Lednick\'y--Lyuboshitz approach (L--L fit, blue bands).
The non-femtoscopic background contributions are represented by the light cyan band. Bottom panels: $n_{\sigma}$ calculated as data-to-fit difference, normalized by the total uncertainty of experimental data.
}
\label{fig:freeFitforpd}
\end{figure*}

The genuine \pd correlation functions for the three centrality intervals considered in this study are presented in Fig.~\ref{fig:freeFitforpd}. 
A flat distribution is assumed for the \pd pairs originating from any of the secondary processes listed in Eq.~\eqref{eq:Cfemto}. 
Therefore, the genuine \pd functions are constrained based on experimental measurements as $C_\text{gen}(\kstar)=(C_\text{femto}(\kstar)-1)/\lambda_\text{gen}+1$. 
The baseline presented in Fig.~\ref{fig:freeFitforpd} as a light cyan line is obtained when considering the event-wise particle correlations due to elliptic flow, which contributes to the femtoscopic range of the \pd correlations in heavy-ion collisions. 
The single-particle azimuthal anisotropies result in a two-particle correlation in relative momentum space, which can be calculated using a data-driven approach. 
The calculation uses a Monte Carlo data resampling method using measured elliptic flow coefficients~\cite{elipio, production77} and momentum distributions of the considered particles. 
The obtained distribution was normalized by the one obtained using the same momenta distributions but with a flat azimuthal anisotropy. 
The resulting baseline shows no significant contribution to the correlation function. 
The vertical bars and gray boxes of the data points represent the statistical and systematic uncertainties, respectively. 
The systematic uncertainties of the \pd correlation functions data points are estimated based on variations of the analysis settings such as the selection of events and particles described in Sec.~\ref{sec:selection}, following the same pattern of settings and methodology as in Ref.~\cite{wiola_pid}.

In order to extract the source size of the \pd system, the genuine \pd correlation function is first calculated using the method developed by the Pisa group~\cite{Viviani2023}. 
This method employs a complete three-body treatment of the \pd system within the Hyperspherical Harmonics (HH) approach~\cite{Kievsky_2008, Marcucci:2019hml} and accounts for strong interactions in \pd by using the AV18 potential~\cite{Av18} for two-nucleon (NN) and the Urbana IX (UIX) potential~\cite{Urbana9} for three-nucleon (NNN) interactions, as well as the Coulomb potential for both the short- and asymptotic-range interactions. 
In the full calculations, all the partial waves up to $J^\pi=\frac{7}{2}^-$ are included providing an excellent description of the \pd correlation function in pp collisions~\cite{Bhawanipd}. 
In the following, the full three-body calculations are referred to as AV18+UIX calculations. Since the AV18+UIX calculations employ nucleons as the degrees of freedom, the relevant source term in this approach is derived under the condition that each nucleon in the three-body system is characterized by the same femtoscopic source radius of the nucleon--nucleon pair, $R_\text{N}$.
The source size $R_{\rm{N}}$ is extracted across the three centrality intervals, from the measured \pd correlations employing the $\chi^2$ minimization method~\cite{kpinPbPb,alice276} on the theoretical correlation functions given by a full three-body approach. 
The values of the source size parameter $R_{\rm{N}}$ are $6.02_{-0.61}^{+1.08}$, $4.04_{-0.58}^{+0.44}$, and $3.14_{-0.36}^{+0.44}$ fm for the 0--10$\%$, 10--30$\%$, and 30--50$\%$ centrality intervals, respectively. 
The modeled correlation function exhibits an excellent agreement with the measured correlation function for all three centrality intervals, as demonstrated in the upper panels of Fig.~\ref{fig:freeFitforpd}. 
Since the NN pairs contribute to forming the p-(p-n) system~\cite{Bhawanipd} at small \kstar, it is reasonable to evenly distribute the total momentum $2k_{\rm{T}}$ of the \pd pair into the three nucleons.
Based on the above definition, the $m_\text{T}$ of a nucleon inside the \pd system can be written as \scalebox{0.85}{$m_{\rm{T}} = \sqrt{\left(2k_{\rm{T}}^{\rm{pd}}/3\right)^2 + m_{\rm{p}}^2}$} and is found to be approximately 1.2~GeV/$c^2$.
The $R_{\rm{N}}$ parameters measured in this study are shown as empty diamond markers in Fig.~\ref{fig:MoneyPlot}. The values of the parameters are in agreement with the proton \mt-scaling measured above. 
This consistency confirms the validity of the \mt-scaling for nucleon pairs that form a \pd system.

Traditionally, the genuine \pd correlation functions can also be modeled using the L--L formalism~\cite{Lednicky:1981su}, which assumes two point-like particles. 
In this method, the deuteron is treated as a single particle rather than the two nucleons being considered individually in the final interaction, which requires scattering parameters as input. 
In this study, these parameters are taken from five different theoretical analyses of \pd scattering data~\cite{Oers1967, Arvieux1974, Black1999, Huttel1983, Kievsky1997}. 
In addition, the source size of the \pd system in the L--L method is determined based on the non-identical pair source expression \mbox{$R_{\rm{ij}}=\sqrt{R^2_{\rm{i}}+R^2_{\rm{j}}}$}~\cite{PhysRevC.81.064906}. 
Here, $R_{\rm{i/j}}$ denotes the single particle femtoscopic source size of a proton and a deuteron for corresponding $m_{\rm{T}}$ of particles that contribute to the \pd $k^*$ range (0, 250)~MeV/$c$. 
In this range, the experimental \pd correlations are fitted using the L--L model with $R_{\rm{pd}}$ as a free parameter by employing the $\chi^2$ minimization method. 
An example of fitting results using representative scattering parameters~\cite{Arvieux1974} is shown as a blue band in Fig.~\ref{fig:freeFitforpd}. 
Similarly to the three-body case, the two-body L--L approach also shows good agreement with the experimental data, unlike the \pd measurement in \pp collisions~\cite{Bhawanipd}. 
The L--L model used for pp collision data with small femtoscopic sources predicts an unphysical peak at very small $k^*$, which is not observed in any data. Removing this artifact requires applying detailed inner corrections or exact solutions of the scattering problem, which accounts for the detailed dynamics of the \pd pair. 
In large systems, such as those in heavy-ion collisions, the average intermediate distances between the considered particles are larger, making the results less sensitive to potential discrepancies caused by approximate solutions in the inner system region~\cite{PhysRevC.111.034903}. 

Furthermore, the deuteron source size is extracted using the expression $R_{\rm{d}} = \sqrt{R^{2}_\text{pd}-R^{2}_\text{p}}$, where $R_\text{p}$ is the single particle femtoscopic source size for the proton, constrained by \mt-scaling of \pP pairs shown in Fig.~\ref{fig:MoneyPlot}. 
The $R_{\rm{d}}$ values are evaluated taking an average of the results obtained by fitting the L--L model with the five different scattering parameters~\cite{Oers1967, Arvieux1974, Black1999, Huttel1983, Kievsky1997} for the strong interaction. 
The values of $R_{\rm{d}}$, reported with total uncertainty, are $4.92_{-0.69}^{+1.02}$, $3.64_{-1.33}^{+1.01}$, and $3.10_{-0.86}^{+1.04}$ fm for the 0--10$\%$, 10--30$\%$, and 30--50$\%$ centrality intervals, respectively. 
The measured femtoscopic source sizes of deuterons are represented in Fig.~\ref{fig:MoneyPlot} by the square markers, corresponding to the \mt of deuterons contributing to low \kstar values, which is approximately 2.4~GeV/$c^2$ for the three centrality intervals.

The extracted single particle source sizes for deuterons are in agreement with \mt-scaling for \pP pairs, suggesting that deuterons, as composite light nuclei, share a common flow velocity with nucleons in \pbpb collisions. 
The observations based on radial and elliptic flow~\cite{production77} bear different natures, thus conclusions from one cannot be directly extrapolated to the other. 
This is because elliptic flow and femtoscopic studies measure the system in different spaces. Therefore, in femtoscopic correlation functions, the effect of elliptic flow is a long-range background beneath the FSI signal. 
On the other hand, the radial flow directly affects the source function, thus the measured radii. 
The single deuteron radii have been previously obtained with larger systematic uncertainties of approximately 50$\%$ in a study of pion-deuteron correlation functions~\cite{wiola_pid}, originating mainly from the imprecise data used to estimate the pion component in the pion-deuteron source. 
The radii measured in the two studies are consistent with the combined statistical and systematic uncertainties. 

The uncertainties of the femtoscopic radii measured in the \pd study are shown in Fig.~\ref{fig:freeFitforpd} with combined statistical and systematic uncertainties. 
They are derived from the 1$\sigma$ range of the distribution of source sizes obtained by resampling the data points of the correlation functions within their systematic and statistical uncertainties, as well as by varying the fitting settings, which include: the upper limit of the fit region 150 or 250~$\mathrm{MeV}/\it{c}$, modification of the $\lambda_{\rm gen}$ factor by $\pm~5\%$, and the baseline obtained from the simulation of the elliptic flow effect or second-order polynomial fit to the normalization region. 
In addition, $R_{\rm{d}}$ accounts for the uncertainty originating from the $R_{\rm{p}}$ source prediction by resampling $R_{\rm{p}}$ using bootstrap over its uncertainties, as well as from the scattering parameters used in the L--L parameterization.

The radii extracted for different \mt values within the same centrality classes for the \pd system are currently compatible within uncertainties for two- and three-body methods. 
However, more precise measurements could enable us to test \mt-scaling for \pd pairs with a level of precision comparable to that already achieved for \pP pairs. 
The ongoing Run 3 at the LHC, where the ALICE detector has been upgraded to handle a rate 100 times larger than that of the Run 2 campaigns, will facilitate these precise measurements~\cite{ALICE_Run34}.

The correlation functions shown in Fig.~\ref{fig:freeFitforpd} demonstrate the repulsive nature of \pd interactions in heavy-ion collisions, which can be accurately described only by including FSI in the model. 
The experimentally observed presence of FSI between the analyzed particles in Pb--Pb collisions, along with the complementary study of p–p collisions~\cite{Bhawanipd}, supports further the notion of an early deuteron formation, as demonstrated in Ref.~\cite{VazquezDoce:2024nye} and theoretically stated in Ref.~\cite{DDCFinNN}.

The fact that both two- and three-body calculations provide an excellent description of the experimental data and that the obtained $R_{\rm N}$ and $R_{\rm d}$ radii values match the nucleon \mt-scaling independently measured for \pP pairs demonstrates the sensitivity of the femtoscopy method, applied to ALICE data for different colliding systems, to the detailed features of the strong interaction. 
Indeed, the previous analysis conducted for pp collisions at 13 TeV demonstrated that the two-body approach fails at small distances due to the short-range features of the interaction, which cannot be adequately described by two-body calculations with the L--L model~\cite{Bhawanipd}. 
However, the three-body calculation provides the same description of the \pd data as the two-body calculation in \pbpb collisions where the inter-particle distances are sufficiently large so that the short-range nature of the interaction no longer affects the correlation observable.

Furthermore, in the two-body case, the obtained radii correspond to the \mt of the deuteron treated as a nucleon-based composite particle, whereas in the three-body case, the measured radii reflect the femtoscopic size of nucleons forming the deuteron at the \mt of a single nucleon. 
The fact that the radii of the deuteron and the nucleon align with the general \mt-scaling trend highlights that the nucleons forming deuterons cannot be distinguished from other nucleons based on this scaling.

\section{Summary}
In conclusion, this work presents the first detailed study of \pP and \pd correlation functions measured in Pb--Pb collisions at $\sqrt{s_{\rm{NN}}}$ = 5.02~TeV by ALICE at the LHC by employing state-of-the-art interaction models. 
The \mt-scaling of proton femtoscopic source radii has been measured for three centrality classes in Pb--Pb collisions, and both two- and three-body calculations have been employed to describe the \pd correlation functions successfully. 
This agreement demonstrates that in Pb--Pb collisions, the three-body approach converges to the two-body ones because of the increasing inter-particle distances. 
The nucleon femtoscopic source radii, obtained from the fits and reported with total uncertainty, range from $2.73^{+0.05}_{-0.05}$~fm in 30–50$\%$ centrality collisions to $5.66^{+0.16}_{-0.16}$~fm in 0–10$\%$ centrality collisions, quantitatively supporting this conclusion.
Moreover, the source values obtained from the two- and three-body fits follow the \mt-scaling of the proton source radii, supporting the hypothesis that the deuteron, as a nucleon-based composite object, shares common flow velocities with protons. 
Since uncertainties are still sizable, more experimental data, particularly from the upcoming Run 3 measurement, will be crucial to verify this hypothesis and improve the precision of these results.


\newenvironment{acknowledgement}{\relax}{\relax}
\begin{acknowledgement}
\section*{Acknowledgements}

The ALICE Collaboration would like to thank all its engineers and technicians for their invaluable contributions to the construction of the experiment and the CERN accelerator teams for the outstanding performance of the LHC complex.
The ALICE Collaboration gratefully acknowledges the resources and support provided by all Grid centres and the Worldwide LHC Computing Grid (WLCG) collaboration.
The ALICE Collaboration acknowledges the following funding agencies for their support in building and running the ALICE detector:
A. I. Alikhanyan National Science Laboratory (Yerevan Physics Institute) Foundation (ANSL), State Committee of Science and World Federation of Scientists (WFS), Armenia;
Austrian Academy of Sciences, Austrian Science Fund (FWF): [M 2467-N36] and Nationalstiftung f\"{u}r Forschung, Technologie und Entwicklung, Austria;
Ministry of Communications and High Technologies, National Nuclear Research Center, Azerbaijan;
Conselho Nacional de Desenvolvimento Cient\'{\i}fico e Tecnol\'{o}gico (CNPq), Financiadora de Estudos e Projetos (Finep), Funda\c{c}\~{a}o de Amparo \`{a} Pesquisa do Estado de S\~{a}o Paulo (FAPESP) and The Sao Paulo Research Foundation  (FAPESP), Brazil;
Bulgarian Ministry of Education and Science, within the National Roadmap for Research Infrastructures 2020-2027 (object CERN), Bulgaria;
Ministry of Education of China (MOEC) , Ministry of Science \& Technology of China (MSTC) and National Natural Science Foundation of China (NSFC), China;
Ministry of Science and Education and Croatian Science Foundation, Croatia;
Centro de Aplicaciones Tecnol\'{o}gicas y Desarrollo Nuclear (CEADEN), Cubaenerg\'{\i}a, Cuba;
Ministry of Education, Youth and Sports of the Czech Republic, Czech Republic;
The Danish Council for Independent Research | Natural Sciences, the VILLUM FONDEN and Danish National Research Foundation (DNRF), Denmark;
Helsinki Institute of Physics (HIP), Finland;
Commissariat \`{a} l'Energie Atomique (CEA) and Institut National de Physique Nucl\'{e}aire et de Physique des Particules (IN2P3) and Centre National de la Recherche Scientifique (CNRS), France;
Bundesministerium f\"{u}r Bildung und Forschung (BMBF) and GSI Helmholtzzentrum f\"{u}r Schwerionenforschung GmbH, Germany;
General Secretariat for Research and Technology, Ministry of Education, Research and Religions, Greece;
National Research, Development and Innovation Office, Hungary;
Department of Atomic Energy Government of India (DAE), Department of Science and Technology, Government of India (DST), University Grants Commission, Government of India (UGC) and Council of Scientific and Industrial Research (CSIR), India;
National Research and Innovation Agency - BRIN, Indonesia;
Istituto Nazionale di Fisica Nucleare (INFN), Italy;
Japanese Ministry of Education, Culture, Sports, Science and Technology (MEXT) and Japan Society for the Promotion of Science (JSPS) KAKENHI, Japan;
Consejo Nacional de Ciencia (CONACYT) y Tecnolog\'{i}a, through Fondo de Cooperaci\'{o}n Internacional en Ciencia y Tecnolog\'{i}a (FONCICYT) and Direcci\'{o}n General de Asuntos del Personal Academico (DGAPA), Mexico;
Nederlandse Organisatie voor Wetenschappelijk Onderzoek (NWO), Netherlands;
The Research Council of Norway, Norway;
Pontificia Universidad Cat\'{o}lica del Per\'{u}, Peru;
Ministry of Science and Higher Education, National Science Centre and WUT ID-UB, Poland;
Korea Institute of Science and Technology Information and National Research Foundation of Korea (NRF), Republic of Korea;
Ministry of Education and Scientific Research, Institute of Atomic Physics, Ministry of Research and Innovation and Institute of Atomic Physics and Universitatea Nationala de Stiinta si Tehnologie Politehnica Bucuresti, Romania;
Ministerstvo skolstva, vyskumu, vyvoja a mladeze SR, Slovakia;
National Research Foundation of South Africa, South Africa;
Swedish Research Council (VR) and Knut \& Alice Wallenberg Foundation (KAW), Sweden;
European Organization for Nuclear Research, Switzerland;
Suranaree University of Technology (SUT), National Science and Technology Development Agency (NSTDA) and National Science, Research and Innovation Fund (NSRF via PMU-B B05F650021), Thailand;
Turkish Energy, Nuclear and Mineral Research Agency (TENMAK), Turkey;
National Academy of  Sciences of Ukraine, Ukraine;
Science and Technology Facilities Council (STFC), United Kingdom;
National Science Foundation of the United States of America (NSF) and United States Department of Energy, Office of Nuclear Physics (DOE NP), United States of America.
In addition, individual groups or members have received support from:
Czech Science Foundation (grant no. 23-07499S), Czech Republic;
FORTE project, reg.\ no.\ CZ.02.01.01/00/22\_008/0004632, Czech Republic, co-funded by the European Union, Czech Republic;
European Research Council (grant no. 950692), European Union;
Deutsche Forschungs Gemeinschaft (DFG, German Research Foundation) ``Neutrinos and Dark Matter in Astro- and Particle Physics'' (grant no. SFB 1258), Germany;
ICSC - National Research Center for High Performance Computing, Big Data and Quantum Computing and FAIR - Future Artificial Intelligence Research, funded by the NextGenerationEU program (Italy).

\end{acknowledgement}

\bibliographystyle{utphys}   
\bibliography{bibliography}

\newpage
\appendix
\section{Scaling parameters}
\label{sec:appendix:param}
The parameters of the fits to the femtoscopic source sizes of protons shown in Fig.~\ref{fig:MoneyPlot} are summarized in Tab.~\ref{tab:powerlawFitmTscaling} and~\ref{tab:linearPara}. Table~\ref{tab:powerlawFitmTscaling} presents the values of the parameters $a$, $b$, and $c$ of the power-law function fit to $R_{\mathrm{p}}$ as a function of transverse mass: $R_{\mathrm{p}} = \mathit{a} + \mathit{b}\cdot\langle m_{\mathrm{T}}\rangle^{\mathit{c}}$. Table~\ref{tab:linearPara} presents the values of the parameters $A$ and $B$ of the linear function fit to the $R_{\mathrm{p}}$ as a function of $\langle\text{dN}_\text{ch}/\text{d}\eta\rangle^{1/3}$: $R_{\text{p}}=\textit{A}\cdot\langle \text{dN}_\text{ch}/\text{d}\eta\rangle^{1/3}+\textit{B}$.

\begin{table}[h!]
    \centering
    \caption{The fitting parameters of the power-law function $R_\text{p} = \it{a} + \it{b}\cdot\langle m_\text{T}\rangle^{\it{c}}$, obtained from the \mt-scaling of the proton source size $R_{\rm{p}}$ shown in Fig.~\ref{fig:MoneyPlot}-(a). 
 }
    \begin{tabular}{c|c|c|c}
      \hline
     Centrality & $a$ & $b$ & $c$ \\ \hline\hline
      0$\mbox{--}$10\%& $3.59\pm0.05$ & $2.78\pm0.04$ & $-1.91\pm0.09$  \\ \hline
      10$\mbox{--}$30\%& $3.38\pm0.02$ & $2.21\pm0.08$ & $-3.16\pm0.14$  \\ \hline
      30$\mbox{--}$50\%& $2.50\pm0.03$ & $1.28\pm0.03$ & $-2.13\pm0.14$  \\ \hline
    \end{tabular}
    \label{tab:powerlawFitmTscaling}
\end{table}

\begin{table}[h!]
    \centering
    \caption{The fitting parameters in the linear function $R_{\mathrm{p}}=\mathit{A}\cdot\langle\mathrm{dN}_{\mathrm{ch}}/\mathrm{d}\eta\rangle^{1/3}+\mathit{B}$,  obtained for the $\langle\text{dN}_\text{ch}/\text{d}\eta\rangle^{1/3}$-scaling of the proton source size $R_{\rm{p}}$ in Fig.~\ref{fig:MoneyPlot}-(b).
    }
  
    \begin{tabular}{c|c|c} 
  \hline
 $m_\text{T}~(\text{GeV}/\textit{c}^{2})$  & $\mathit{A}$ & $\mathit{B}$ \\ \hline\hline

  $[1.0,1.3)$ & $0.496\pm 0.002$ & $-0.39\pm 0.02$ \\ \hline
  $[1.3,1.4)$& $0.428\pm 0.001$ &  $-0.05\pm 0.01$ \\ \hline
  $[1.4,1.5)$& $0.392\pm 0.001$ & $0.14\pm 0.01$ \\ \hline
  $[1.5,1.6)$& $0.390\pm 0.001$ & $0.07\pm 0.01$ \\ \hline
  $[1.6,1.7)$& $0.388\pm 0.001$ &  $0.02\pm 0.01$ \\ \hline
  $[1.7,1.8)$& $0.355\pm 0.001$ & $0.25\pm 0.01$ \\ \hline
  $[1.8,2.0)$& $0.341\pm 0.001$ &  $0.26\pm 0.01$ \\ \hline 
  $[2.0,3.2)$&  $0.314\pm 0.001$ & $0.39\pm 0.01$ \\ \hline
  
\end{tabular}

    \label{tab:linearPara}
\end{table}
\newpage


%
%

\section{The ALICE Collaboration}
\label{app:collab}
\begin{flushleft} 
\small

S.~Acharya\,\orcidlink{0000-0002-9213-5329}\,$^{\rm 51}$, 
G.~Aglieri Rinella\,\orcidlink{0000-0002-9611-3696}\,$^{\rm 33}$, 
L.~Aglietta\,\orcidlink{0009-0003-0763-6802}\,$^{\rm 24}$, 
M.~Agnello\,\orcidlink{0000-0002-0760-5075}\,$^{\rm 30}$, 
N.~Agrawal\,\orcidlink{0000-0003-0348-9836}\,$^{\rm 25}$, 
Z.~Ahammed\,\orcidlink{0000-0001-5241-7412}\,$^{\rm 135}$, 
S.~Ahmad\,\orcidlink{0000-0003-0497-5705}\,$^{\rm 15}$, 
S.U.~Ahn\,\orcidlink{0000-0001-8847-489X}\,$^{\rm 73}$, 
I.~Ahuja\,\orcidlink{0000-0002-4417-1392}\,$^{\rm 37}$, 
Z.~Akbar$^{\rm 83}$
% ...

\section*{Affiliation Notes}

$^{\rm I}$ Deceased\\
$^{\rm II}$ Also at: Max-Planck-Institut fur Physik, Munich, Germany\\
$^{\rm III}$ Also at: Italian National Agency for New Technologies, Energy and Sustainable Economic Development (ENEA), Bologna, Italy\\
$^{\rm IV}$ Also at: Instituto de Fisica da Universidade de Sao Paulo\\
$^{\rm V}$ Also at: Dipartimento DET del Politecnico di Torino, Turin, Italy\\
$^{\rm VI}$ Also at: Department of Applied Physics, Aligarh Muslim University, Aligarh, India\\
$^{\rm VII}$ Also at: Institute of Theoretical Physics, University of Wroclaw, Poland\\
$^{\rm VIII}$ Also at: Facultad de Ciencias, Universidad Nacional Aut\'{o}noma de M\'{e}xico, Mexico City, Mexico\\

\section*{Collaboration Institutes}

$^{1}$ A.I. Alikhanyan National Science Laboratory (Yerevan Physics Institute) Foundation, Yerevan, Armenia\\
$^{2}$ AGH University of Krakow, Cracow, Poland\\
$^{3}$ Bogolyubov Institute for Theoretical Physics, National Academy of Sciences of Ukraine, Kiev, Ukraine\\
$^{4}$ Bose Institute, Department of Physics  and Centre for Astroparticle Physics and Space Science (CAPSS), Kolkata, India\\
$^{5}$ California Polytechnic State University, San Luis Obispo, California, United States\\
$^{6}$ Central China Normal University, Wuhan, China\\
$^{7}$ Centro de Aplicaciones Tecnol\'{o}gicas y Desarrollo Nuclear (CEADEN), Havana, Cuba\\
$^{8}$ Centro de Investigaci\'{o}n y de Estudios Avanzados (CINVESTAV), Mexico City and M\'{e}rida, Mexico\\
$^{9}$ Chicago State University, Chicago, Illinois, United States\\
$^{10}$ China Nuclear Data Center, China Institute of Atomic Energy, Beijing, China\\
$^{11}$ China University of Geosciences, Wuhan, China\\
$^{12}$ Chungbuk National University, Cheongju, Republic of Korea\\
$^{13}$ Comenius University Bratislava, Faculty of Mathematics, Physics and Informatics, Bratislava, Slovak Republic\\
$^{14}$ Creighton University, Omaha, Nebraska, United States\\
$^{15}$ Department of Physics, Aligarh Muslim University, Aligarh, India\\
$^{16}$ Department of Physics, Pusan National University, Pusan, Republic of Korea\\
$^{17}$ Department of Physics, Sejong University, Seoul, Republic of Korea\\
$^{18}$ Department of Physics, University of California, Berkeley, California, United States\\
$^{19}$ Department of Physics, University of Oslo, Oslo, Norway\\
$^{20}$ Department of Physics and Technology, University of Bergen, Bergen, Norway\\
$^{21}$ Dipartimento di Fisica, Universit\`{a} di Pavia, Pavia, Italy\\
$^{22}$ Dipartimento di Fisica dell'Universit\`{a} and Sezione INFN, Cagliari, Italy\\
$^{23}$ Dipartimento di Fisica dell'Universit\`{a} and Sezione INFN, Trieste, Italy\\
$^{24}$ Dipartimento di Fisica dell'Universit\`{a} and Sezione INFN, Turin, Italy\\
$^{25}$ Dipartimento di Fisica e Astronomia dell'Universit\`{a} and Sezione INFN, Bologna, Italy\\
$^{26}$ Dipartimento di Fisica e Astronomia dell'Universit\`{a} and Sezione INFN, Catania, Italy\\
$^{27}$ Dipartimento di Fisica e Astronomia dell'Universit\`{a} and Sezione INFN, Padova, Italy\\
$^{28}$ Dipartimento di Fisica dell'Universit\`{a} and Sezione INFN, Pisa, Italy\\
$^{29}$ Dipartimento di Fisica `E.R.~Caianiello' dell'Universit\`{a} and Gruppo Collegato INFN, Salerno, Italy\\
$^{30}$ Dipartimento DISAT del Politecnico and Sezione INFN, Turin, Italy\\
$^{31}$ Dipartimento di Scienze MIFT, Universit\`{a} di Messina, Messina, Italy\\
$^{32}$ Dipartimento Interateneo di Fisica `M.~Merlin' and Sezione INFN, Bari, Italy\\
$^{33}$ European Organization for Nuclear Research (CERN), Geneva, Switzerland\\
$^{34}$ Faculty of Electrical Engineering, Mechanical Engineering and Naval Architecture, University of Split, Split, Croatia\\
$^{35}$ Faculty of Nuclear Sciences and Physical Engineering, Czech Technical University in Prague, Prague, Czech Republic\\
$^{36}$ Faculty of Physics, Sofia University, Sofia, Bulgaria\\
$^{37}$ Faculty of Science, P.J.~\v{S}af\'{a}rik University, Ko\v{s}ice, Slovak Republic\\
$^{38}$ Faculty of Technology, Environmental and Social Sciences, Bergen, Norway\\
$^{39}$ Frankfurt Institute for Advanced Studies, Johann Wolfgang Goethe-Universit\"{a}t Frankfurt, Frankfurt, Germany\\
$^{40}$ Fudan University, Shanghai, China\\
$^{41}$ Gangneung-Wonju National University, Gangneung, Republic of Korea\\
$^{42}$ Gauhati University, Department of Physics, Guwahati, India\\
$^{43}$ Helmholtz-Institut f\"{u}r Strahlen- und Kernphysik, Rheinische Friedrich-Wilhelms-Universit\"{a}t Bonn, Bonn, Germany\\
$^{44}$ Helsinki Institute of Physics (HIP), Helsinki, Finland\\
$^{45}$ High Energy Physics Group,  Universidad Aut\'{o}noma de Puebla, Puebla, Mexico\\
$^{46}$ Horia Hulubei National Institute of Physics and Nuclear Engineering, Bucharest, Romania\\
$^{47}$ HUN-REN Wigner Research Centre for Physics, Budapest, Hungary\\
$^{48}$ Indian Institute of Technology Bombay (IIT), Mumbai, India\\
$^{49}$ Indian Institute of Technology Indore, Indore, India\\
$^{50}$ INFN, Laboratori Nazionali di Frascati, Frascati, Italy\\
$^{51}$ INFN, Sezione di Bari, Bari, Italy\\
$^{52}$ INFN, Sezione di Bologna, Bologna, Italy\\
$^{53}$ INFN, Sezione di Cagliari, Cagliari, Italy\\
$^{54}$ INFN, Sezione di Catania, Catania, Italy\\
$^{55}$ INFN, Sezione di Padova, Padova, Italy\\
$^{56}$ INFN, Sezione di Pavia, Pavia, Italy\\
$^{57}$ INFN, Sezione di Pisa, Pisa, Italy\\
$^{58}$ INFN, Sezione di Torino, Turin, Italy\\
$^{59}$ INFN, Sezione di Trieste, Trieste, Italy\\
$^{60}$ Inha University, Incheon, Republic of Korea\\
$^{61}$ Institute for Gravitational and Subatomic Physics (GRASP), Utrecht University/Nikhef, Utrecht, Netherlands\\
$^{62}$ Institute of Experimental Physics, Slovak Academy of Sciences, Ko\v{s}ice, Slovak Republic\\
$^{63}$ Institute of Physics, Homi Bhabha National Institute, Bhubaneswar, India\\
$^{64}$ Institute of Physics of the Czech Academy of Sciences, Prague, Czech Republic\\
$^{65}$ Institute of Space Science (ISS), Bucharest, Romania\\
$^{66}$ Institut f\"{u}r Kernphysik, Johann Wolfgang Goethe-Universit\"{a}t Frankfurt, Frankfurt, Germany\\
$^{67}$ Instituto de Ciencias Nucleares, Universidad Nacional Aut\'{o}noma de M\'{e}xico, Mexico City, Mexico\\
$^{68}$ Instituto de F\'{i}sica, Universidade Federal do Rio Grande do Sul (UFRGS), Porto Alegre, Brazil\\
$^{69}$ Instituto de F\'{\i}sica, Universidad Nacional Aut\'{o}noma de M\'{e}xico, Mexico City, Mexico\\
$^{70}$ iThemba LABS, National Research Foundation, Somerset West, South Africa\\
$^{71}$ Jeonbuk National University, Jeonju, Republic of Korea\\
$^{72}$ Johann-Wolfgang-Goethe Universit\"{a}t Frankfurt Institut f\"{u}r Informatik, Fachbereich Informatik und Mathematik, Frankfurt, Germany\\
$^{73}$ Korea Institute of Science and Technology Information, Daejeon, Republic of Korea\\
$^{74}$ Laboratoire de Physique Subatomique et de Cosmologie, Universit\'{e} Grenoble-Alpes, CNRS-IN2P3, Grenoble, France\\
$^{75}$ Lawrence Berkeley National Laboratory, Berkeley, California, United States\\
$^{76}$ Lund University Department of Physics, Division of Particle Physics, Lund, Sweden\\
$^{77}$ Nagasaki Institute of Applied Science, Nagasaki, Japan\\
$^{78}$ Nara Women{'}s University (NWU), Nara, Japan\\
$^{79}$ National and Kapodistrian University of Athens, School of Science, Department of Physics , Athens, Greece\\
$^{80}$ National Centre for Nuclear Research, Warsaw, Poland\\
$^{81}$ National Institute of Science Education and Research, Homi Bhabha National Institute, Jatni, India\\
$^{82}$ National Nuclear Research Center, Baku, Azerbaijan\\
$^{83}$ National Research and Innovation Agency - BRIN, Jakarta, Indonesia\\
$^{84}$ Niels Bohr Institute, University of Copenhagen, Copenhagen, Denmark\\
$^{85}$ Nikhef, National institute for subatomic physics, Amsterdam, Netherlands\\
$^{86}$ Nuclear Physics Group, STFC Daresbury Laboratory, Daresbury, United Kingdom\\
$^{87}$ Nuclear Physics Institute of the Czech Academy of Sciences, Husinec-\v{R}e\v{z}, Czech Republic\\
$^{88}$ Oak Ridge National Laboratory, Oak Ridge, Tennessee, United States\\
$^{89}$ Ohio State University, Columbus, Ohio, United States\\
$^{90}$ Physics department, Faculty of science, University of Zagreb, Zagreb, Croatia\\
$^{91}$ Physics Department, Panjab University, Chandigarh, India\\
$^{92}$ Physics Department, University of Jammu, Jammu, India\\
$^{93}$ Physics Program and International Institute for Sustainability with Knotted Chiral Meta Matter (WPI-SKCM$^{2}$), Hiroshima University, Hiroshima, Japan\\
$^{94}$ Physikalisches Institut, Eberhard-Karls-Universit\"{a}t T\"{u}bingen, T\"{u}bingen, Germany\\
$^{95}$ Physikalisches Institut, Ruprecht-Karls-Universit\"{a}t Heidelberg, Heidelberg, Germany\\
$^{96}$ Physik Department, Technische Universit\"{a}t M\"{u}nchen, Munich, Germany\\
$^{97}$ Politecnico di Bari and Sezione INFN, Bari, Italy\\
$^{98}$ Research Division and ExtreMe Matter Institute EMMI, GSI Helmholtzzentrum f\"ur Schwerionenforschung GmbH, Darmstadt, Germany\\
$^{99}$ Saga University, Saga, Japan\\
$^{100}$ Saha Institute of Nuclear Physics, Homi Bhabha National Institute, Kolkata, India\\
$^{101}$ School of Physics and Astronomy, University of Birmingham, Birmingham, United Kingdom\\
$^{102}$ Secci\'{o}n F\'{\i}sica, Departamento de Ciencias, Pontificia Universidad Cat\'{o}lica del Per\'{u}, Lima, Peru\\
$^{103}$ Stefan Meyer Institut f\"{u}r Subatomare Physik (SMI), Vienna, Austria\\
$^{104}$ SUBATECH, IMT Atlantique, Nantes Universit\'{e}, CNRS-IN2P3, Nantes, France\\
$^{105}$ Sungkyunkwan University, Suwon City, Republic of Korea\\
$^{106}$ Suranaree University of Technology, Nakhon Ratchasima, Thailand\\
$^{107}$ Technical University of Ko\v{s}ice, Ko\v{s}ice, Slovak Republic\\
$^{108}$ The Henryk Niewodniczanski Institute of Nuclear Physics, Polish Academy of Sciences, Cracow, Poland\\
$^{109}$ The University of Texas at Austin, Austin, Texas, United States\\
$^{110}$ Universidad Aut\'{o}noma de Sinaloa, Culiac\'{a}n, Mexico\\
$^{111}$ Universidade de S\~{a}o Paulo (USP), S\~{a}o Paulo, Brazil\\
$^{112}$ Universidade Estadual de Campinas (UNICAMP), Campinas, Brazil\\
$^{113}$ Universidade Federal do ABC, Santo Andre, Brazil\\
$^{114}$ Universitatea Nationala de Stiinta si Tehnologie Politehnica Bucuresti, Bucharest, Romania\\
$^{115}$ University of Derby, Derby, United Kingdom\\
$^{116}$ University of Houston, Houston, Texas, United States\\
$^{117}$ University of Jyv\"{a}skyl\"{a}, Jyv\"{a}skyl\"{a}, Finland\\
$^{118}$ University of Kansas, Lawrence, Kansas, United States\\
$^{119}$ University of Liverpool, Liverpool, United Kingdom\\
$^{120}$ University of Science and Technology of China, Hefei, China\\
$^{121}$ University of South-Eastern Norway, Kongsberg, Norway\\
$^{122}$ University of Tennessee, Knoxville, Tennessee, United States\\
$^{123}$ University of the Witwatersrand, Johannesburg, South Africa\\
$^{124}$ University of Tokyo, Tokyo, Japan\\
$^{125}$ University of Tsukuba, Tsukuba, Japan\\
$^{126}$ Universit\"{a}t M\"{u}nster, Institut f\"{u}r Kernphysik, M\"{u}nster, Germany\\
$^{127}$ Universit\'{e} Clermont Auvergne, CNRS/IN2P3, LPC, Clermont-Ferrand, France\\
$^{128}$ Universit\'{e} de Lyon, CNRS/IN2P3, Institut de Physique des 2 Infinis de Lyon, Lyon, France\\
$^{129}$ Universit\'{e} de Strasbourg, CNRS, IPHC UMR 7178, F-67000 Strasbourg, France, Strasbourg, France\\
$^{130}$ Universit\'{e} Paris-Saclay, Centre d'Etudes de Saclay (CEA), IRFU, D\'{e}partment de Physique Nucl\'{e}aire (DPhN), Saclay, France\\
$^{131}$ Universit\'{e}  Paris-Saclay, CNRS/IN2P3, IJCLab, Orsay, France\\
$^{132}$ Universit\`{a} degli Studi di Foggia, Foggia, Italy\\
$^{133}$ Universit\`{a} del Piemonte Orientale, Vercelli, Italy\\
$^{134}$ Universit\`{a} di Brescia, Brescia, Italy\\
$^{135}$ Variable Energy Cyclotron Centre, Homi Bhabha National Institute, Kolkata, India\\
$^{136}$ Warsaw University of Technology, Warsaw, Poland\\
$^{137}$ Wayne State University, Detroit, Michigan, United States\\
$^{138}$ Yale University, New Haven, Connecticut, United States\\
$^{139}$ Yildiz Technical University, Istanbul, Turkey\\
$^{140}$ Yonsei University, Seoul, Republic of Korea\\
$^{141}$ Affiliated with an institute formerly covered by a cooperation agreement with CERN\\
$^{142}$ Affiliated with an international laboratory covered by a cooperation agreement with CERN.\\

\end{flushleft} 

\end{document}